# Mobility variations in mono- and multi-layer graphene films

Kosuke Nagashio*, Tomonori Nishimura, Koji Kita and Akira Toriumi
Department of Materials Engineering, The University of Tokyo, Tokyo 113-8656, JAPAN
*nagashio@material.t.u-tokyo.ac.jp

**Abstract**

The electric properties of mono- and multi-layer graphene films were systematically studied with the layer number determined by their optical contrast. The current modulation increased monotonically with a decrease in the layer number due to the reduction of the interlayer scattering. Carrier mobility in the monolayer was significantly greater than that in the multilayer due to linear dispersion relation. On the other hand, in the monolayer, carrier transport was extremely sensitive to charged impurity density due to the reduction in screening effect, which causes larger mobility variation. Reduction of the charged impurity density is thus key for high mobility.

Graphene, a single layer of graphite, has attracted both scientific and technological interests as a new two-dimensional electron gas system with extremely high mobility of ~10,000 $cm^2$/Vs at room temperature.[1)] The density of electrons and holes can be modulated by the gate electric field due to a low density of states of the graphene. Although there are many reports of graphene on $SiO_2$, one problem is that the mobility varies widely from 2,000 to 20,000 $cm^2$/Vs[2,3)], even with two-probe measurement results excluded. Another problem is that the layer number dependence is still unclear, since current research focuses mainly on mono- and bi-layer graphene films.[4)] Monolayer and multilayer graphene films possess a linear dispersion and parabolic ones with the band overlapping, respectively.[5)] Monolayer graphene film is clearly distinguished from multilayer films by 2D band around 2700 $cm^{-1}$ in the Raman spectrum.[6)] The layer number dependence of the spectra for multilayer graphene films, however, is not decisive, and the spectrum shape and position are also strongly affected by doping.[7)] Atomic force microscopy is not suitable to determine the layer number due to an uncertain dead space between multilayer graphene films and $SiO_2$.[8)] Therefore, in order to systematically investigate electrical properties as a function of the layer number, a simple but reliable layer number determination method is required.

In the present study, we first show that the combination of optical microscopy and Raman spectroscopy can determine the layer number, and that electrical transport properties (specifically, mobility and Dirac point shift) can then be reported as a function of the layer number.

Monolayer and multilayer graphene films were transferred onto 90-nm $SiO_2$/$p^+$-Si substrates via the micromechanical cleavage of Kish graphite. After a suitable graphene film had been selected under the optical microscope, electron-beam lithography was utilized in order to pattern electrical contacts on the graphene. Thermal evaporation of Cr/Au (5/25 nm) was followed by lift-off in warm acetone. To remove the resist residual, graphene devices were annealed in a $H_2$-Ar mixture at 300 °C for 1 hour[8)] and then electrical measurements were performed in vacuum. Using a Hall-bar-type electrode configuration (see the inset in Fig. 4(a)), Hall measurements were also performed.

Figure 1 shows graphene contrast as a function of $SiO_2$ thickness calculated using the Fresnel equation, assuming a trilayer model of graphene, $SiO_2$ and Si. This is based upon calculation by Blake et al.[9)], who first showed that the visibility of graphene is determined by the wavelength and $SiO_2$ thickness from the difference in reflected light intensities, R, with and without graphene as follows,

$$contrast = \frac{R(w/o \cdot g) - R(w \cdot g)}{R(w/o \cdot g)}. \qquad [1]$$

In the present calculation, contrasts for the wavelengths between 450 and 750 nm are superimposed to represent the total contrast for visible light, including intensity dependence on the wavelength for the halogen lamp and the neutral colour balance filter. The inset shows both R(w/o·g) and R(w·g) superimposed for visible light wavelengths.

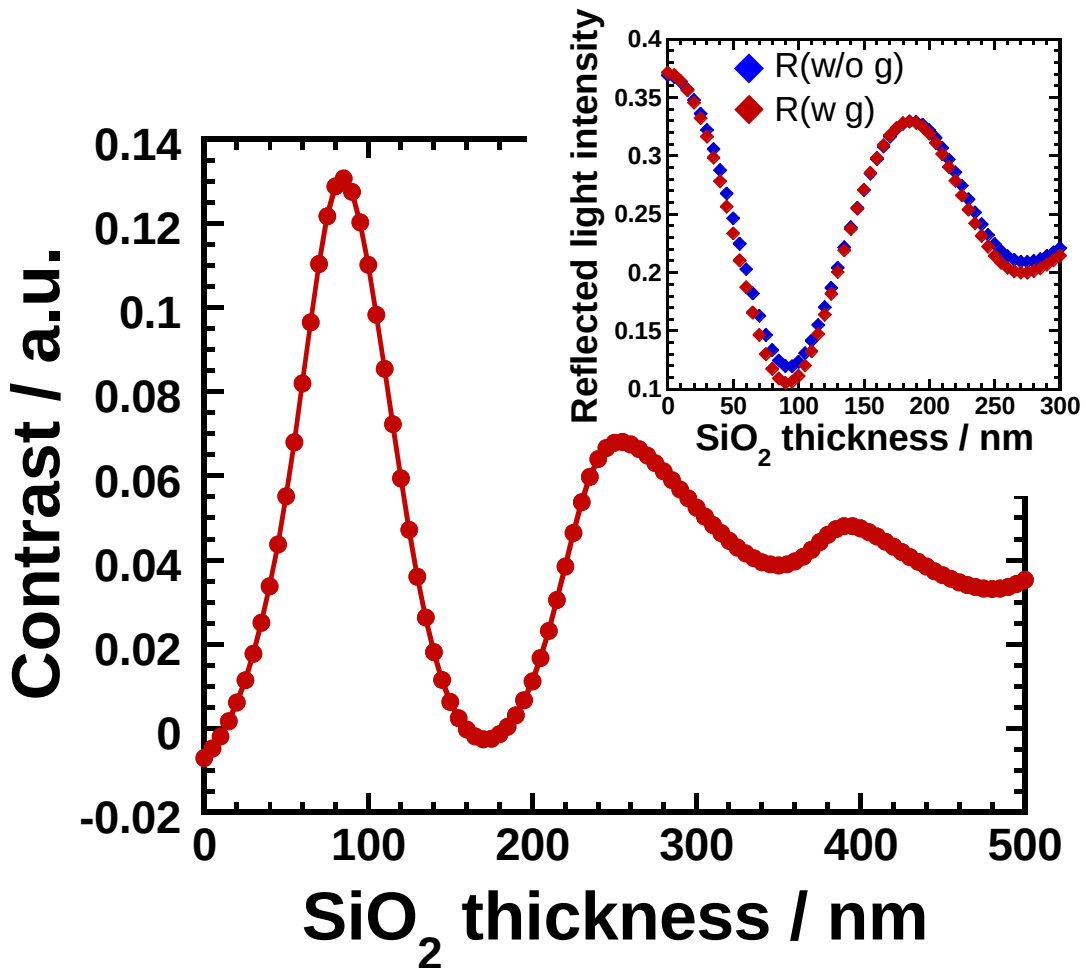

**Figure 1** Contrast of graphene as a function of $SiO_2$ thickness as calculated using the Fresnel equation assuming a trilayer model consisting of graphene, $SiO_2$ and Si. Inset shows R(w/o·g) and R(w·g) superimposed for visible light wavelengths. A small but finite difference between R(w/o·g) and R(w·g) makes the contrast observable.

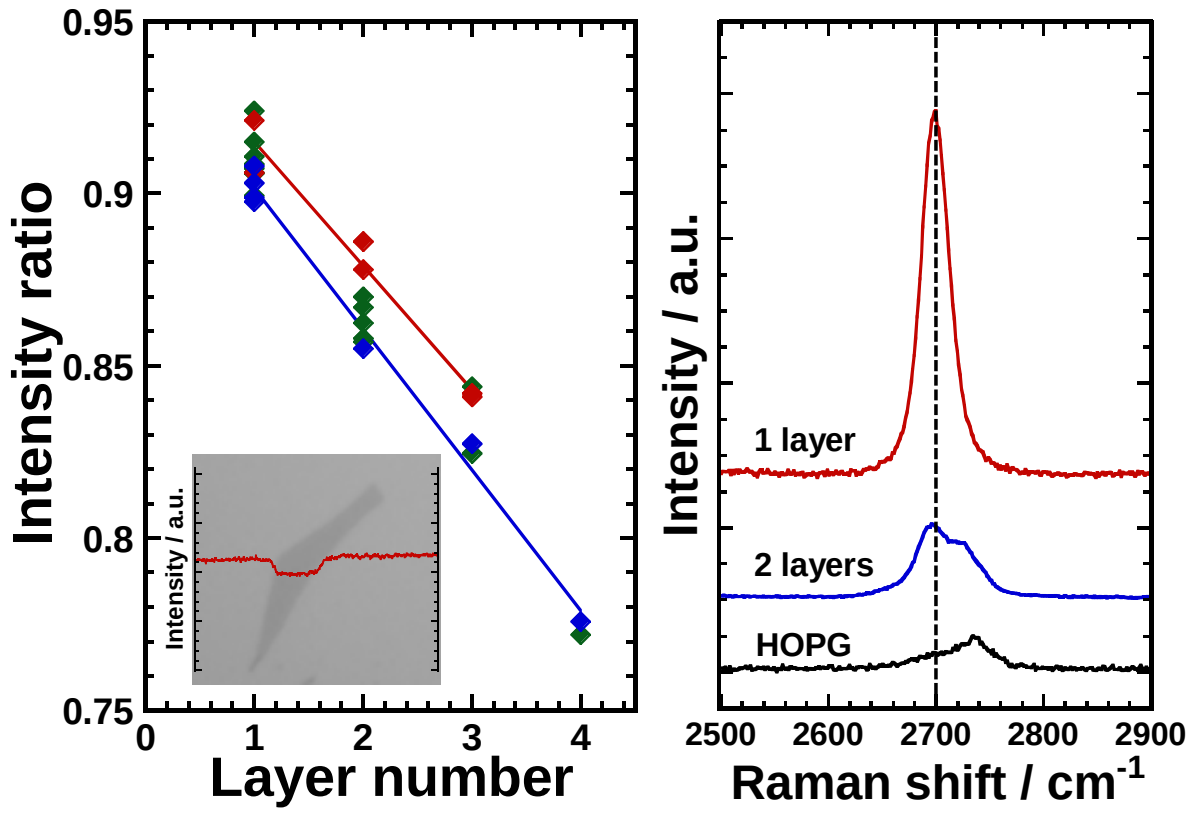

(a) (b)

**Figure 2** (a) Intensity ratio of graphite thin films on 90nm $SiO_2$ (squares) as a function of layer number. The inset shows an optical micrograph and intensity profile of monolayer graphene. (b) Raman spectra of 2D band for highly oriented pyrolytic graphite (HOPG), bilayer, and monolayer graphene films.

It is found that the periodicity of the contrast is visible due to interference periodicity in $SiO_2$ for R(w/o·g). From the comparison of the inset and main figures, graphene is seen on the $SiO_2$ thickness with the lowest R(w/o·g), where the existence of a 0.335-nm-thick graphene film gives a small but measurable difference (the difference between blue and red squares in the inset). The most striking result is that the contrast of graphene on 90 nm $SiO_2$ is higher than that on 300 nm, which is calculated by considering all wavelengths of visible light. Regarding transconductance enhancement in the back gate structure, thinner $SiO_2$ is preferable. The insulator thickness can be further reduced to ~35 nm by replacing $SiO_2$ with $TiO_2$, which has a larger refractive index of 2.7.

The inset of Figure 2(a) shows an optical micrograph of the monolayer graphene observed on a 90 nm $SiO_2$/Si substrate. Many mono- and multi-layer graphene films whose contrasts depend on the layer number can be easily found on a $SiO_2$/Si wafer. After conversion of the color to gray scale, the intensity ratio of graphene to $SiO_2$ was analyzed based on a line profile. As reference for the intensity ratio, only the monolayer was confirmed by a 488 nm Raman spectrum of the 2D band, as shown in Fig. 2(b), since a single peak in the Raman spectrum is direct evidence for monolayer graphene.[6)] The intensity ratio is plotted as a function of the layer number in Fig. 2(a). Even for monolayer graphene, the intensity ratios are scattered and some of them are close to those observed for the bilayer graphene film. However, when the intensity-ratio data for mono- and multilayer graphene films on the same $SiO_2$/Si wafer are selected as shown by red and blue squares with lines (two examples), the intensity ratio is clearly separated and is linearly dependent on the layer number. Therefore, this simple intensity-ratio analysis using the gray-scale images can be used to determine the layer number. Although Ni et al. suggested optical layer counting without the Raman spectrum,[10)] the present experiment has shown that the monolayer graphene reference as determined by the Raman spectrum is critical even for 90 nm $SiO_2$ with a higher contrast than 300 nm $SiO_2$ because of the uncertain dead space between the graphene and $SiO_2$. This optical method is applicable for graphene multilayer films thinner than ~4 nm (~10 layers) because the intensity ratio increases with the layer number on the contrary.

Figure 3 shows (a) sheet resistivity and (b) conductivity as a function of the gate voltage $V_g$ for mono- and multi-layer graphene films with different layer number. All data were obtained by four-probe measurement, since graphene is not degraded significantly by the deposition of metal electrodes unlike the carbon nanotube (CNT) case.[11)] It should be noted that a back gate voltage of 30 V for 90 nm $SiO_2$ is almost equivalent to 100 V for 300 nm $SiO_2$. Sheet resistivity monotonically increases with a decrease in layer number; for the monolayer, the resistivity curve drastically changes and has the smallest full-width at half-maximum but with almost the same resistivity at the Dirac point as that of the bilayer graphene film.

An important finding is that the Dirac point shifts to a negative voltage as the layer number increases. The Dirac point is very sensitive to charged impurities such as resist residue, atmospheric $H_2O$ and so on.[1)] Therefore, the resist residue was carefully removed via $H_2$/Ar annealing[8)] and the effect of atmospheric $H_2O$ was avoided by performing electric measurements in vacuum. The Dirac point in the present experiments moved to the same negative voltages after $H_2$/Ar annealing. These results strongly suggest that the Dirac point shifts with the layer number and is intrinsic to graphene films. Here, if the work function changes gradually from monolayer graphene to graphite with a increase in layer number, the charge transfer may take place between the metal electrodes and the graphene film. This would cause the Fermi level to shift due to the relatively low density states of mono- and multi-layer graphene films. In fact, it has been reported that the work function increased from mono- to bi-layer graphene films for graphene films on SiC.[12)] However, qualitative study is further

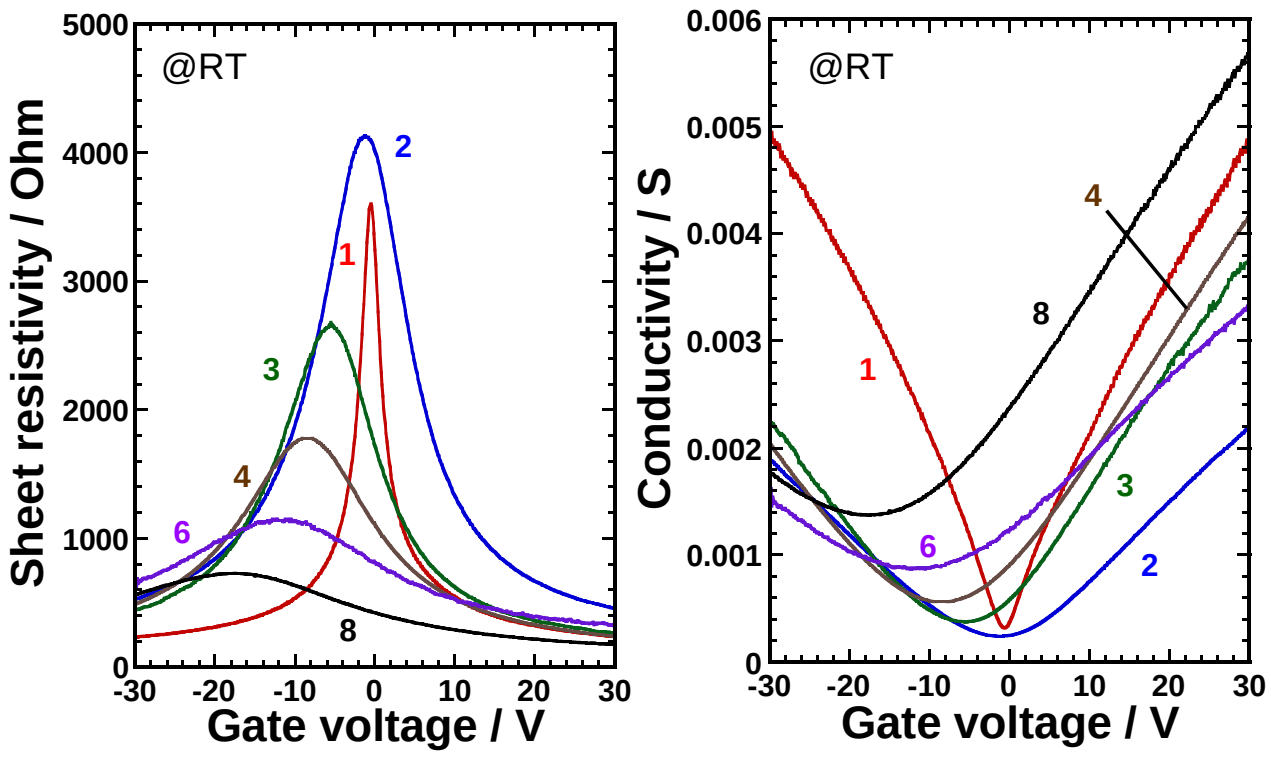

(a) (b)

**Figure 3** (a) sheet resistivity and (b) conductivity as a function of gate voltage for graphene films with different layer number.

required to elucidate the origin of the Dirac point shift.

Figure 3(b) shows the linear relationship between conductivity and gate voltage. The conductivity-Vg curve for the monolayer is sharper than that of others. This is attributable to the much lower density state of the graphene monolayer compared to graphene multilayers due to the difference in the dispersion relationship. Moreover, only for the monolayer graphene, it was observed that the conductivity curve changed from a linear to a sublinear curve with increasing mobility (not shown), as reported previously.[13)]

Figure 4(a) shows the mobility for mono- and multilayer graphene films with different layer number. Since it was found that the carrier density, as determined by the Hall measurement, approximately agreed with the surface charge density capacitively induced by the back gate, the mobility was evaluated by $\mu=1/en\rho$ where the carrier density n was obtained from $n=\varepsilon_{ox}/d_{ox}(V_g-V_{Dirac})/e$, and $\varepsilon_{ox}$ and $d_{ox}$ are the permittivity and thickness of $SiO_2$, respectively. The mobilities of 2, 3 and 4 graphene layers are all roughly 3000 $cm^2/Vs$ . The highest mobility achieved, 8200 $cm^2/Vs$, was in the monolayer graphene at room temperature, but there is a great deal of scatter in the data. When the temperature was reduced to 20 K, the mobility increased to ~12,000 $cm^2/Vs$ for the monolayer graphene, while it did not change for multilayer graphene films. The mean free path for the highest mobility of the monolayer graphene at 20 K was calculated to be ~350 nm by $\lambda=v_F\tau$ where $v_F$ is the Fermi velocity of ~$10^6$ m/s and $\tau=\hbar\sigma(\pi/n)^{1/2}/(e^2v_F)$. It was much shorter than the present device size (~8 μm), which suggests that the transport in the monolayer graphene is not ballistic, but is rather diffusive.

It has been reported that the conductivity of the monolayer graphene is limited by charged impurity scattering in the case of the linear relationship between conductivity and carrier density. [14)] According to the following equation,[14)]

$$\sigma = 20\frac{e^2}{h}\frac{n}{n_{imp}}, \qquad [2]$$

the mobility can be evaluated as a function of the charged impurity density $n_{imp}$ as shown in Fig. 4 (b), where experimental results in Fig. 4 (a) are also plotted. This result suggests that a large variation of the charged impurity density in the range of 5.8-$15\times10^{11}$ $cm^{-2}$ may result in wide scattering of the measured mobility.

With decreasing layer number, the current modulation may be enhanced due to the reduction of the interlayer scattering. On the other hand, when the layer number is decreased from the bi- to mono-layer graphene, the mobility drastically increased due to the inherent change from the quadric to linear dispersion relationships. The screening length in multilayer graphene films is reported, theoretically, to be 0.5 nm[15)] and 0.7 nm[16)], and experimentally to be 1.2

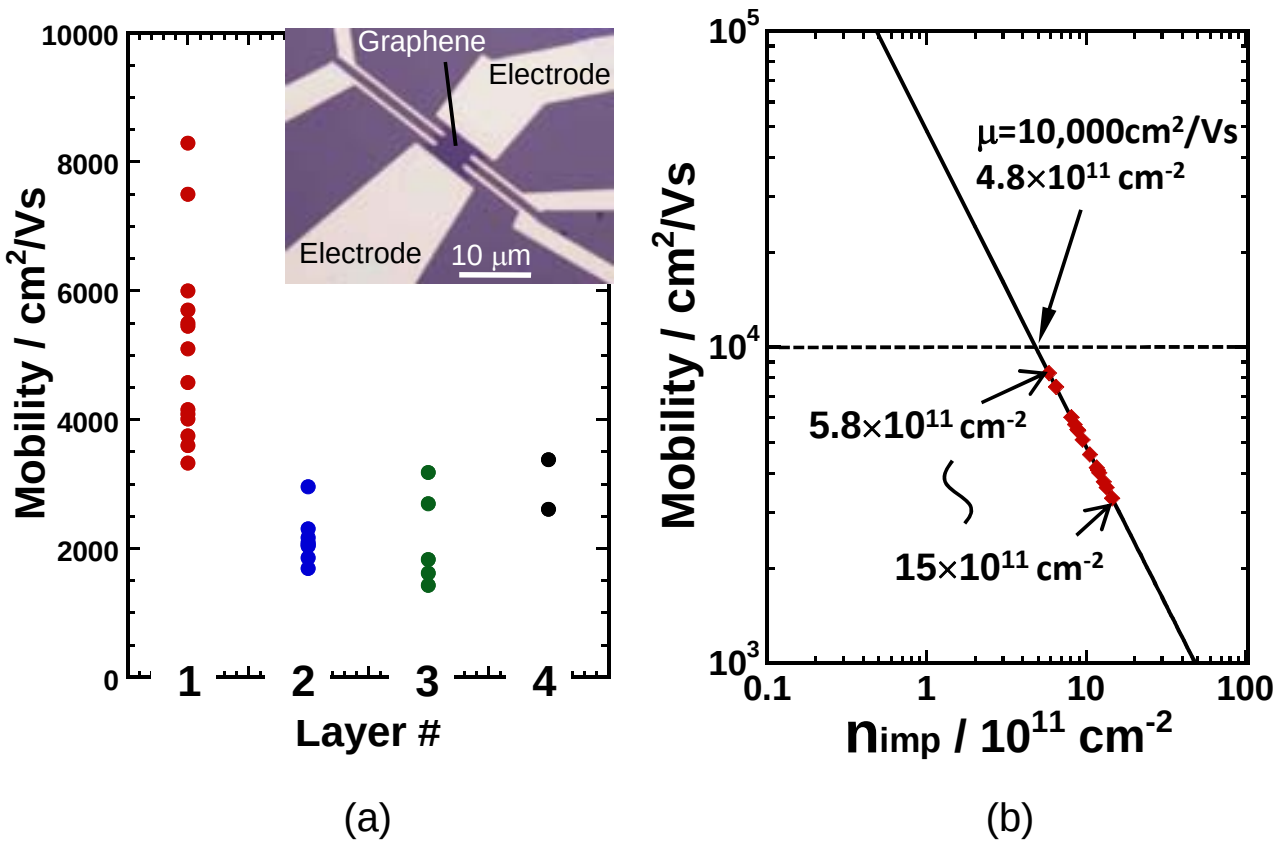

**Figure 4** (a) Mobility variation for mono- and multi-layer graphene films with different layer number. Inset shows the optical micrograph of a typical graphene FET with the Hall-bar structure. (b) Mobility vs. charged impurity density calculated using equation [2]. The solid symbols show the experimental results in (a).

nm.[17)] The charges induced by gate voltage are mainly located within one or two layers considering the inter-graphene layer distance of 0.335 nm. Electric transport of the monolayer graphene should be quite sensitive to charged impurities as the scattering origin due to the screening effect reduction. It is concluded that the large mobility variation for the monolayer graphene is caused by variations in the charged impurity density. If the charged impurity density can be reduced to below $4.8\times10^{11}$ $cm^{-2}$, the mobility will exceed 10,000 $cm^2/Vs$ at room temperature.

In summary, the number of layers in multilayer graphene films was optically determined using a simple and reliable method using a monolayer graphene reference. Electric characterization for graphene films with well-determined layer numbers revealed that the electric field effectively modulated the carrier transport for the monolayer graphene due to the linear dispersion. The mobility became very sensitive to charged impurities as a scattering origin due to a decrease in the screening effect, which resulted in large mobility variations. Reduction of the charged impurity density at the graphene/$SiO_2$ interface is key to achieve monolayer graphene with a mobility greater than 10,000 $cm^2/Vs$ at room temperature.

**Acknowledgements**

We would like to thank Prof. Maruyama, University of Tokyo, for use of the Raman spectroscope. Kish graphite used in this study was kindly provided by Covalent Materials Co.